\documentstyle[sprocl,epsfig]{article}

\newcommand{\be}{\beta}

\newcommand{\lsim}%

\def\be{\begin{equation}}
\def\ee{\end{equation}}
\def\bea{\begin{eqnarray}}
\def\eea{\end{eqnarray}}
\begin{document}

\title{SUSY LOOP CONTRIBUTIONS TO THE $W$-PAIR PRODUCTION IN $e^+e^-$ 
       COLLISIONS}

\author{KAORU HAGIWARA}

\address{Theory Group, KEK, Tukuba, Ibaraki 305-0801, Japan}

\author{SHINYA KANEMURA}

\address{Department of Physics, Osaka University, Toyonaka, Osaka
         560-0043, Japan}

\author{YOSHIAKI UMEDA}

\address{Institute of Physics, National Chiao Tung Univeristy, Hsinchu
         300, Taiwan}


\maketitle\abstracts{
We discuss one-loop contributions of supersymmetry (SUSY) particles 
to the process $e^+e^- \to W^+W^-$ in the minimal supersymmetric 
standard model. The calculation is
tested by using the sum rule among the form factors, which is 
based on the BRS invariance.
The overall normalization factor of the amplitudes is tested 
by using the decoupling property of the SUSY particles 
in the large mass limit.
The correction due to the one-loop effect of the sfermions can be 
of a few times $\pm 0.1$ \% level, 
whereas that of the charginos and neutralinos can be order $1$\%. 
We also study the CP-violating effects in 
the chargino and neutralino sector.
}

%***********************************************************************

%\section{Introduction}

We present the results of our study on 
the one-loop super-partner particle contributions 
to the $e^+(k,\tau)e^-(\overline{k},\overline{\tau})
\rightarrow W^+(p,\lambda) W^- (\overline{k},\overline{\lambda})$ 
in the MSSM \cite{brs}, where $k$ and $p$ represent the momenta, 
and $\tau$ and $\lambda$ do the helicities. 
We here concentrate on the one-loop effects
of sfermions \cite{sf}, charginos and neutralinos \cite{ino}.

The helicity amplitudes can be written by using 16 basis tensors 
$T_i^{\mu\alpha\beta}$ as
\vspace*{-3mm}
\begin{equation}
\label{amp-eeww}
{M}^{\lambda \overline{\lambda}}_\tau
= \sum_{i=1}^{16} F_{i,\tau}(s,t)\, j_\mu(k,\overline{k},\tau) 
T_i^{\mu\alpha\beta} \epsilon_\alpha(p,\lambda)^\ast 
\epsilon_\beta(\overline{p},\overline{\lambda})^\ast  \;,
\vspace*{-3mm}
\end{equation}
where $j_{\mu}$ is the electron current, and 
$\epsilon_{\alpha}$ is the polarization vectors for the $W$ boson.
The physical process is described by the first 9 of the 16 form factors 
$F_{i,\tau}(s,t)$ ($i=1$-$16$).
In order to test the physical form factors $F_{i,\tau}(s,t)$($i=1$-$9$) 
by using the BRS sum rule \cite{brs}, 
we have to calculate unphysical form factors 
$F_{i,\tau}(s,t)$ ($i=10$-$16$) together
with the form factors $H_{i,\tau}(s,t)$ ($i=1$-$4$) of 
helicity amplitudes for $e^+(k,\tau)e^-(\overline{k},\overline{\tau})
\rightarrow \omega^+(p) W^- (\overline{k},\overline{\lambda})$
($\omega^+$: Nambu-Goldstone boson) by using basis tensors 
$S^{\mu\alpha}_i$ as
\vspace*{-3mm}
\begin{equation}
 {M}^\lambda_\tau (e^+e^-\rightarrow \omega^+ W^-)
 =    i \sum_{i=1}^{4} H_{i,\tau}(s,t)\, j_\mu(k,\overline{k},\tau) 
S_i^{\mu\alpha} \epsilon_\alpha(p,\lambda)^\ast \;.
\label{amp-eewx}
\vspace*{-3mm}
\end{equation}
We employ the $\overline{\rm MS}$ scheme for the one-loop 
calculation \cite{sf,ino}.
  
%\section{Test of the one-loop calculation}

One difficulty in loop-level calculations is to determine reliability 
of the results. This is especially so in our process in which a subtle 
gauge cancellation takes place among diagrams at each level of 
perturbation. 
In order to obtain solid results, we test 
our calculation by using the sum rules among the form factor 
due to the BRS invariance\cite{brs,sf,ino}. 
In addition, using the decoupling property of the SUSY particles in the 
large mass limit, we can test the overall normalization 
of the amplitudes\cite{sf,ino}. 

\begin{table}[t]
\begin{center}
{\small
\begin{tabular}{l|rrr}
{\sf $\!\!\!\!$ First 2 generations} $\!\!\!\!\!\!$ 
& Case 1 & Case 2 & Case 3   \\ \hline
\multicolumn{4}{l}{Input parameters }           \\ \hline
$m_{\tilde{Q}}$= 
$m_{\tilde{U}}$=  
$m_{\tilde{D}}$ & 300 & 500 &1000 \\
$A_{\tilde{f}}^{\sf eff}$&   0 &   0 &   0  \\ 
\hline  \hline
\end{tabular} 
\caption{Cases without squark mixing for Fig.~1(left).}
\begin{tabular}{l|rrr}
{\sf $\tilde{t}$-$\tilde{b}$ sector:}  & Case 1 & Case 2 & Case 3   \\ \hline
\multicolumn{4}{l}{Input parameters }           \\ \hline
$m_{\tilde{Q}}$= 
$m_{\tilde{U}}$=  
$m_{\tilde{D}}$ & 300 & 400 &500 \\
$A_{\tilde{f}}^{\sf eff}$& 625 & 1025 & 1539 \\ \hline
\hline  
\multicolumn{4}{l}{Output parameters} \\ \hline
$m_{\tilde{t}_1}$ & 100 & 100 &  100 \\
$m_{\tilde{t}_2}$ & 478 & 607 & 741 \\
$\cos\theta_{\tilde{t}}$ &0.708&0.708&0.707 \\
\hline\hline 
\end{tabular}
\caption{Maximal stop-mixing cases for Fig.~1(right).}
}
\end{center}
\end{table}

     \begin{figure}[t]
     \begin{center}
     \begin{tabular}{cc}
     \mbox{\epsfig{file=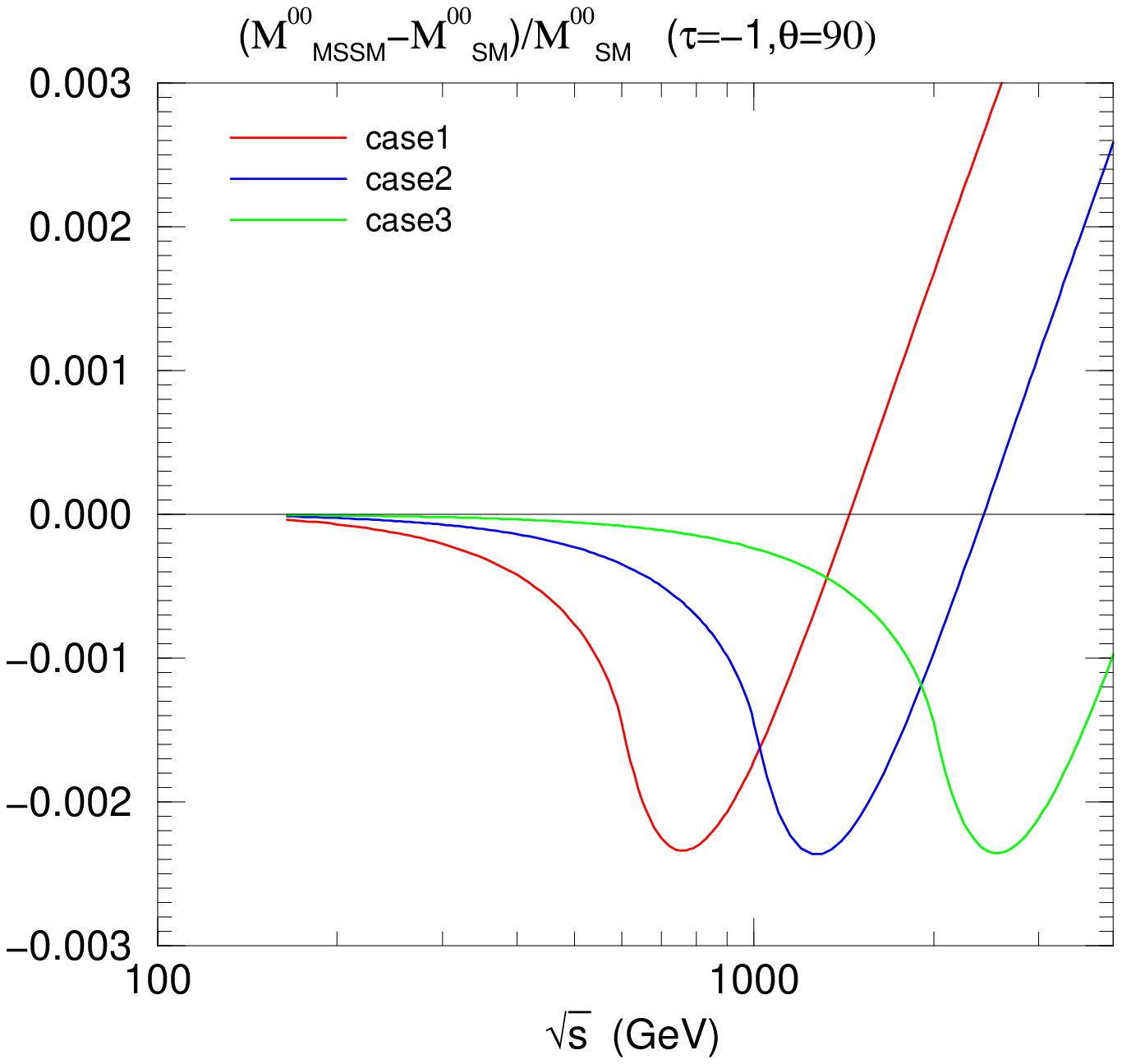,width=4.2cm}} &
     \mbox{\epsfig{file=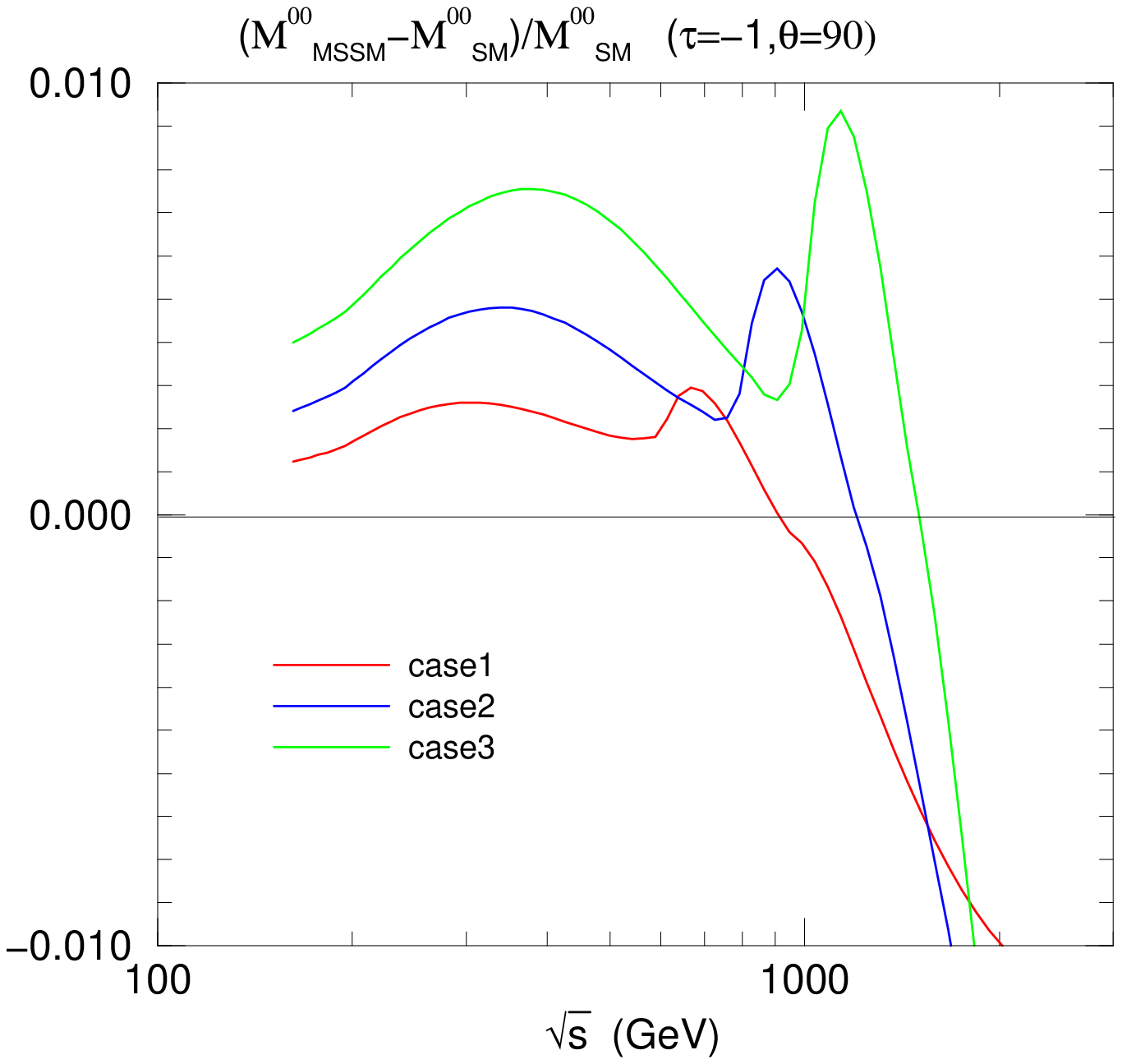,width=4.2cm}}
     \end{tabular}
     \caption{Squark contributions for non-LR-mixing (left) and for 
       maximal LR-mixing (right).}
     \label{fig:wwzz}
     \end{center}
     \end{figure}

%\section{Results}

%\subsection{Sfermion one-loop effects}

Among the helicity amplitudes for $e^+e^- \to W^+W^-$, 
the SUSY loop contribution is the largest in the 00 helicity amplitude.
In Fig.~1 (left), the squark one-loop effects 
on the $00$ helicity amplitude are shown for parameter sets in Table 1. 
The corrections to the SM prediction are negative and 
the behavior is rather simple. 
It is found that the corrections 
to the SM prediction are at most a few times 0.1\%.
In Fig.~1(right), effects of the third generation squarks 
with large stop-mixing are shown.  The parameters defined in  
Table~2 are chosen so as to be the maximal mixing 
with the mixing angle 45$^\circ$. The corrections are positive. 
Larger effects appear for larger values of $A^{\sf eff}_f$. 
However, it turns out that such enhancement due to the 
mixing is strongly constrained by the precision data \cite{sf}.  
The cases for large corrections (case 2, case 3 in Table 2) 
stay outside the 99\% contour of the allowed region.
Consequently, only smaller corrections than a few times 0.1\% 
are allowed. 

\begin{table}[t]
{\small
\begin{center}
\begin{tabular}{l|rrrrr}
   & A & B & C & D & E\\ \hline
\multicolumn{5}{l}{Parameter}           \\ \hline
$\mu$ (GeV)     & +120 & +145 & +400 & +1000 & +130 \\
$M_2$ (GeV)     & 541 & 242 & 125 & 115  & 158 \\
$\varphi_{1}$  &0&   0 &   0 &    0 & $\frac{2}{3}\pi$ \\
$\varphi_{\mu}$&0&   0 &   0 &    0 & $\frac{2}{3}\pi$ \\
%\hline  
%\multicolumn{5}{l}{Mass spectra (GeV)} \\ \hline
%$m_{\widetilde{\chi}^-_1}$  & 110 & 110 & 110 & 110 & 110\\
%$m_{\widetilde{\chi}^-_2}$  & 555 & 283 & 420 &1007 & 207\\
%$m_{\widetilde{\chi}^0_1}$  &  99 &  81 &  60 &  57 &  75\\
%$m_{\widetilde{\chi}^0_2}$  & 123 & 150 & 111 & 110 & 105\\
%$m_{\widetilde{\chi}^0_3}$  & 285 & 150 & 403 &1002 & 154\\
%$m_{\widetilde{\chi}^0_4}$  & 555 & 285 & 422 &1007 & 205\\
\hline\hline
\end{tabular}\label{tab:set1}
\end{center}
\caption{Parameter sets for Figs.~2 and 3. }
}
\end{table}

%\subsection{Chargino and neutralino one-loop effects}

Next, we discuss the chargino and neutralino loop effects.
We here assume the GUT relation $M_1 = 5M_2\hat{s}^2/3\hat{c}^2$, 
the light chargino mass $m_{\widetilde{\chi}_1^-}$ = 110GeV, 
and the scattering angle $\theta = 90^\circ$.
In Fig.~2(a), we show $\mu$ parameter dependences for
$\tan\beta = 3$ and $50$. 
The corrections are about 0.7\% at $|\mu| = 1000$ GeV for $\tau =-1$,
whereas they remain to be small for $\tau =+1$ as around $0.1\%$ level.
In Fig.~2(b), $\sqrt{s}$ dependences  
for four sets of parameter (sets A to D of Table~3)
are shown. The lightest chargino mass is fixed to be 
$m_{\tilde{\chi}^-}^{}=110$ GeV, and $\tan\beta=3$ is assumed for 
all cases. 
A sharp peak can be seen at $\sqrt{s}= 220$GeV for each curve, 
which corresponds to the threshold for the pair production of 
the lightest chargino.
The corrections are negative for $\tau =+1$.
%

%\subsection{$CP$-violating effects}

     \begin{figure}[t]
     \begin{center}
     \mbox{\epsfig{file=4f.epsi,width=6.3cm}}
     \caption{One-loop chargino and neutralino effects on the 
          $M^{00}_\tau$ amplitude.}
     \label{fig:wwzz}
     \mbox{\epsfig{file=7f.epsi,width=6.3cm}}
     \caption{The effect of the CP violating phases.}
     \label{fig:wwzz}
     \end{center}
     \end{figure}

There are new sources for the $CP$-violating phases in the MSSM.
In the chargino and neutralino sector, it arises from
the $\mu$-term, $\mu e^{i \varphi_{\mu}}$, and from 
the gaugino mass parameters $M_1 e^{i\varphi_1}$ 
when we take the phase of $M_2$ to be zero by rephasing. 
We find that large $CP$-violating phases 
in chargino and neutralino
sectors are possible without contradicting the EDM constraints.
In Figs.~3(a) and 3(b), the real part (solid curve) 
and the imaginary part (dotted curve) of $f_4^Z$, $f_6^\gamma$,
and $f_6^Z$ are shown as a function of $\sqrt{s}$
for the parameters of set E in Table~3 \cite{ino}.
These form factors take their maximum or minimum at around
$\varphi_{\mu}=\varphi_{1}=2/3\pi$ or $4/3\pi$.
The $CP$-violating effects of $f_4$ and $f_6$ may be measured 
through the difference of the helicity amplitudes 
$M^{\pm 0}$ and $M^{\mp 0}$.
The $CP$-violating effects in the helicity amplitudes can be
of the order 0.1\%. 
%
%\section{Conclusion}

In conclusion, the sfermion one-loop contributions 
to the ${00}$  helicity amplitude 
are at most a few times $0.1$\%, while 
the effects due to charginos and neutralinos can be 
of the order of 1\%. We also found that 
the $CP$-violating effects on the helicity amplitudes 
$M^{\pm 0}$ and $M^{\mp 0}$ 
can be of the order of 0.1\%.

\end{document}